\documentclass[12pt]{article}

\usepackage{etex}
\usepackage{amsmath}
\usepackage{amsfonts}
\usepackage{amssymb}
\usepackage{amsthm}
\usepackage{amssymb}
\usepackage[all]{xy}
\usepackage{graphicx}
\usepackage{indentfirst}
\usepackage{diagram}
\usepackage{makecell}
\usepackage[figuresright]{rotating}
\usepackage{pdflscape}
\usepackage{thmtools}
\usepackage{pst-node}
\usepackage{tikz-cd}
\usepackage{setspace}
\usepackage{pgf}        
\usepackage{hyperref}
\usepackage{url}
\usepackage{tikz}
\usetikzlibrary{shapes.geometric, arrows}

\usepackage{epigraph}

\numberwithin{equation}{section}

\makeatletter
\renewcommand{\@biblabel}[1]{#1\hfill \hspace{-0.2cm}}
\makeatother

\renewcommand{\vec}{\boldsymbol}

\usepackage{yfonts}
\newfont{\gothic}{eufm10 at 12pt}

\newtheorem{theorem}{Theorem}[section]

\theoremstyle{definition}

\newtheorem{proposition}[theorem]{Proposition}

\numberwithin{equation}{section}
\author{Timothy F. Power\footnote{Timothy.Power@dal.ca} \,   and Roman G. Smirnov\footnote{e-mail:  Roman.Smirnov@dal.ca} \\[0.5cm]Department of
  Mathematics and Statistics\\
Dalhousie University\\ Halifax, Nova Scotia, Canada
  B3H~3J5}
\begin{document}
\title{Rethinking Growth: An Extension of the Solow-Swan Model}

\maketitle

\begin{abstract}
 The aggregate Cobb-Douglas production function stands as a central element in the renowned Solow-Swan model in economics, providing a crucial theoretical framework for comprehending the determinants of economic growth.  This model not only guides policymakers and economists but also influences their decisions, fostering sustainable and inclusive development. In this study, we utilize a one-input version of a new generalization of the Cobb-Douglas production function proposed recently, thereby extending the Solow-Swan model to incorporate energy production as a factor. We offer a rationale for this extension and conduct a comprehensive analysis employing advanced mathematical tools to explore solutions to this new model. This approach allows us to effectively integrate environmental considerations related to energy production into economic growth strategies, fostering long-term sustainability. 
\end{abstract}

\section{Introduction}\label{sec1}

The objective of our paper is to extend the widely recognized Solow-Swan model  \cite{RS1956, TS1956}  in light of recent developments in economic growth theory (see, for example,  Cangiotti and Sensi \cite{CS2022} and the relevant references therein for recent advances and generalizations). Additional developments and references  can be found    in \cite{BMK2022}, \cite{CET2012}, \cite{NCJ2019}, and \cite{DK2012}. 

As is well-known, the model offers valuable insights into long-term economic growth and the underlying factors driving it. It is widely acknowledged that an aggregate  production function $Y= f(K,L)$, normally defined  to be a function of $K$ (capital) and $L$ (labor), as well as the differential equation describing capital accumulation form the core elements of the Solow-Swan model. Depending on the context, the economy in question is assumed to either generate aggregate output or produce a single good that can be allocated to consumption or investment. The production output is governed by the Cobb-Douglas production function \cite{CD1928}:

\begin{equation}
\label{cd}
Y = A K^{\alpha}L^{\beta},
\end{equation} 
where $A$ represents total factor productivity, and the output elasticities $0<\alpha<1$ and $0<\beta<1$ are typically constrained by the condition 
\begin{equation}
\label{one}
\alpha + \beta = 1.
\end{equation}
This condition guarantees that the function $Y$ in (\ref{cd}) displays constant returns to scale, i.e., it is a homogeneous function of degree one. The Solow-Swan model posits that economic growth stems from capital accumulation, growth in the labor force, and exogenous technological progress. The standard aggregate production function, generalizing (\ref{cd}), is specified as follows:
\begin{equation}
\label{1}
Y(t)=K(t)^{\alpha}(A(t)L(t))^{1-\alpha},
\end{equation}
where $Y$ (production) is generated by $L$ (labor), $K$ (capital), and $A$ (labor-augmenting technology or ``knowledge"). It is also assumed the functions $L(t)$ and $A(t)$ grow exponentially: $L(t) = L(0)e^{nt}$ and $A(t) = A(0)e^{gt}$.  At the same time, the stock of capital depreciates at a constant rate, while only a fraction of production is invested, thus leading to the following differential equation describing capital accumulation:
\begin{equation}
\label{2}
\dot{K}(t)= sY(t) - \delta K(t),
\end{equation}
where $\dot{K}(t) = \mbox{d}K/\mbox{d}t$,  the constant $0<s<1$ is the investment rate while the constant $\delta$ represents the depreciation of capital. 

Since the production function $Y = f(K, AL)$ given by (\ref{1}) enjoys constant returns to scale, a new variable $y(t)$ representing the output output per effective unit of labor $y(t) = \frac{Y(t)}{A(t)L(t)}$ can be introduced, with respect to which the function (\ref{1}) reduces to 
$$y(t) = k(t)^{\alpha},$$
where $k(t) = \frac{K(t)}{A(t)L(t)}$ represents the capital intensity. Accordingly, in terms of the variables $y(t)$ and $k(t)$ the differential equation (\ref{2}) reduces to 
\begin{equation}
\label{2a}
\dot{k}(t) = sk(t)^{\alpha} - (n+ g + \delta)k(t). 
\end{equation} 
We note that the differential equation (\ref{2a}) implies that the capital intensity $k(t)$ converges to a steady-state value determined by the condition $sk(t)^{\alpha} - (n+ g + \delta)k(t)=0$.  See \cite{BSM2003, DR2012} for more details. 

In our methodology, we introduce key modifications to the classical Solow-Swan model. First, we replace the Cobb-Douglas production function \cite{CD1928} given by (\ref{cd}) -- which depends on capital and labor—with an energy-centric formulation, offering a novel perspective on economic growth. This modification is justified for two primary reasons.

On the one hand, the Cobb-Douglas function no longer aligns with contemporary economic development data (see, e.g., \cite{SW2019} and references therein). On the other hand, extensive research explores the relationship between GDP and energy production. While most studies employ statistical methods, some have adopted analytical approaches using differential equations, dynamical systems, and even Hamiltonian formalism (see \cite{CS2021, DS2011, DK2012} and references therein).

Furthermore, we argue that theories explaining growth solely through energy supply --  while neglecting information, knowledge, and institutions --  remain incomplete. Conversely, we recognize that these very factors (information, knowledge, and institutions) are themselves fundamentally dependent on energy availability. As David Stern notes in \cite{DS2011}: ``Because thermodynamics implies energy is essential to all economic production, criticism of mainstream economic growth models that ignore energy is legitimate."

This perspective is further supported by Beaudreau \cite{BCB2016}, who examines why energy's role was historically marginalized in growth theories and analyzes the consequences of this oversight for production function theory.

In this view, we conceptualize the production function as a single input factor function, with energy production serving as the independent variable. Subsequently, we employ the single-factor version of the production function introduced in \cite{SW2019} as a generalized adaptation of the Cobb-Douglas function within a logistic growth model to generalize the celebrated Solow-Swan model. 

\section{Mathematical model design and assumptions}

In this section, we highlight the key features that differentiate our model from the classical Solow-Swan framework and discuss the motivations behind its development. First, we justify the departure from the Cobb-Douglas functional form traditionally used in the Solow-Swan model. Instead, we adopt a more general production function, which encompasses the Cobb-Douglas function as a special case.

As outlined above, the Solow-Swan model is centered on  the Cobb-Douglas production function (\ref{cd}-\ref{1}), which was studied by Charles Cobb and Paul Douglas in their seminal 1928 paper \cite{CD1928} by linking it to a specific data set. However, it is important to note that the function had previously been introduced and examined by Knut Wicksell, Philip Wicksteed, and L\'{e}on Walras (see \cite{H1997} and the relevant references therein). Cobb and Douglas in \cite{CD1928} established its legitimacy through a combination of economic reasoning, statistical methods, and empirical data. Specifically, they demonstrated that assigning the values $\alpha = 0.25$, $\beta = 0.75$, and $A = 1.01$ to the respective parameters in (\ref{cd}) resulted in a production function with constant returns to scale, yielding a precise fit to the data describing the growth of the American economy between 1899 and 1922. 

In recent years, several researchers, including the second author (RGS) of this paper, have extended the approach by Cobb and Douglas employed in \cite{CD1928} by incorporating recent advances in the theory of data-driven dynamical systems into their study of the Cobb-Douglas function and its properties \cite{SWW2022, SW2024}.  Explicitly, instead of fitting the function given by (\ref{cd}) to the above mentioned dataset, they used R programming to fit the system of differential equations 
\begin{equation}
\label{sys}
\begin{array}{rcl}
\dot{K}(t) & = & b_1K(t), \\[0.3cm]
\quad \dot{L}(t) &=& b_2 L(t), \\ [0.3cm]
\quad \dot{Y}(t) &=& b_3Y(t),
\end{array}
\end{equation} 
describing the exponential growth in  capital, labor, and production. Thus, it was shown with the aid of R programming  that the solutions 
\begin{equation}
\label{sol}
K(t) = K_0e^{b_1t}, \quad L(t) = L_0e^{b_2t}, \quad Y(t) = Y_0e^{b_3t} 
\end{equation}
were compatible with the data studied in \cite{CD1928} (see also \cite{SW2021}) for the following numerical values: 
\begin{equation}
\label{data1}
\begin{array}{lll}

b_1=0.06472564, &  \ln K_0=4.61213588 & \mbox{(capital)},\\ [0.3cm]

b_2=0.02549605, & \ln L_0=4.66953290 & \mbox{(labor)}, \\[0.3cm]

b_3=0.03592651, & \ln Y_0 =4.66415363 & \mbox{(production)}.
\end{array}
\end{equation}
Note that in the above $b_2<b_3<b_1$.  On the other hand, using the method moving frames, alas at a rudimentary level (see \cite{CMS11, CMS17, HMS05} for more details and references), we can combine the functions  (\ref{sol}) as follows: 
\begin{equation}
\label{inv}
K^{a_1}L^{a_2}Y^{a_3} = K_0^{a_1}L_0^{a_2}Y_0^{a_3} e^{(a_1b_1+a_2b_2+a_3b_3)t},
\end{equation}
where $a_1,a_2, a_3 \in \mathbb{R}\setminus \{0\}$ are some parameters. Clearly, the function (\ref{inv}) is a time-independent invariant of the one-parameter Lie group action defined by (\ref{sol}) iff the {\em orthogonality condition}
\begin{equation}
\label{con}
a_1b_1 + a_2 b_2 + a_3b_3 = 0
\end{equation} 
holds. Obviously, the condition (\ref{con}) is equivalent to $\vec{a}\cdot \vec{b} = 0$, where $\vec{a} = <a_1, a_2, a_3>$, $\vec{b} = <b_1, b_2, b_3>$, hence the name.  Next, solving  in (\ref{inv}) for $Y$ under the orthogonality condition (\ref{con}), we obtain
\begin{equation}
\label{cd1}
Y = AK^{-\frac{a_1}{a_3}}L^{-\frac{a_2}{a_3}},
\end{equation}
where $A = (K_0^{a_1}L_0^{a_2}Y_0^{a_3})^{\frac{1}{a_3}}$. Defining 
\begin{equation}
\label{ab}
\alpha := -\frac{a_1}{a_3}, \quad \beta := - \frac{a_2}{a_3},
\end{equation}
we rewrite the condition (\ref{con}) as 
\begin{equation}
\label{con1}
\alpha b_1 +\beta b_2 - b_3 =0
\end{equation} 
and arrive at the following one-parameter family of the Cobb-Douglas functions of the form (\ref{cd}): 
\begin{equation}
\label{inv1}
Y = A K^{\alpha} L^{\frac{b_3}{b_2} - \alpha\frac{b_1}{b_2}}, \quad \alpha>0.
\end{equation} 
The key question now is whether the Cobb-Douglas function with constant returns to scale belongs to the one-parameter family (\ref{inv1}). By solving the system of linear equations in $\alpha$ and $\beta$ (\ref{one}) and (\ref{con1}), we arrive at the formulas
\begin{equation}
\label{ab1}
\alpha = \frac{b_2-b_3}{b_2-b_1}, \quad \beta= \frac{b_3-b_1}{b_2-b_1},
\end{equation}
which hold true for $0<\alpha, \beta<1$, provided  $b_2<b_3<b_1$. Going in the opposite direction is analogous. Therefore, we have proven the following
\begin{proposition}
{\em The exponential model (\ref{sys}) admits a one-parameter family of Cobb-Douglas-type functions (\ref{inv1}) as a time-independent invariant if and only if the orthogonality condition (\ref{con}) is satisfied. Moreover, the Cobb-Douglas function exhibiting constant returns to scale is a member of this family if and only if $b_2<b_3<b_1$. }
\end{proposition}
See \cite{SWW2022, SW2024} for more details. The formula (\ref{ab1}) was originally derived using the bi-Hamiltonian approach \cite{PS05}   in \cite{SW2021}. Notably, the values of $b_1$, $b_2$, and $b_3$ retrieved from the data set studied by Cobb and Douglas in 1928 \cite{CD1928} (\ref{data1}) satisfy this inequality. This is precisely why the Cobb-Douglas function $Y = 1.01 K^{0.25}L^{0.75}$ was shown to be a good fit  for the given data in \cite{CD1928}.  Next, substituting the values for $b_1$, $b_2$, and $b_3$ given by (\ref{data1}) into (\ref{ab1}), we get the following values for $\alpha$ and $\beta$: 
\begin{equation}
\label{alphabeta-1}
\alpha =0.2658824627, \quad \beta=0.7341175376. 
\end{equation}
Predictably, they are nearly identical to those obtained in \cite{CD1928} during the process of fitting the function (\ref{cd}) to the data directly. In light of this, we can construct infinitely many production functions of the Cobb-Douglas type, given by (\ref{inv1}), that fit the data studied by Cobb and Douglas in \cite{CD1928} -- see \cite{SW2024} for examples. This observation highlights that in order to specify the appropriate Cobb-Douglas function within the one-parameter family (\ref{inv1}) that provides a good fit to a given dataset, one must seek and incorporate additional information beyond the time series for capital, labor, and production. For example, labor share data, as was shown in  \cite{SW2024}, can serve as a crucial complement in this selection process. In fact, that is what Cobb and Douglas themselves (implicitly) used in \cite{CD1928} to justify the choice for $\alpha$ and $\beta$, assuming (\ref{one}): ``A further interesting comparison is afforded by the studies of the National Bureau of Economic Research into the proportion of the manufacturing product which went to labor during the decade 1909-1918. They found that wages  and salaries  formed on the average 74 per cent of the total value added by manufacturers  during those years. We have found in our formula that when we attribute to labor 75 per cent of the product, we get a close consilience to the actual normal course of production."

However, it is important to note that if the growth of capital, labor, and production does not follow an exponential trend, no function of the Cobb-Douglas type (\ref{inv1}) can be fitted to the corresponding data with  a good accuracy. We argue that this is the primary reason why the Cobb-Douglas production function is no longer broadly compatible with current economic data (see \cite{SW2019,SW2024}  for more details and references).
Accordingly, we modify the exponential model (\ref{sys}) as follows, introducing a logistic model under the assumption that capital, labor, and production now grow logistically rather than exponentially.
\begin{equation}
\label{sys1}
\begin{array}{rcl}
\dot{K}(t) &=& b_1K(t)\left(1-\frac{K(t)}{N_K}\right), \\[0.3cm]

\dot{L}(t) &=& b_2 L(t)\left(1-\frac{L(t)}{N_L}\right), \\[0.3cm]

 \dot{Y}(t) &=& b_3Y(t)\left(1-\frac{Y(t)}{N_Y}\right).
 \end{array}
\end{equation} 
The constants $N_K$, $N_L$, and $N_Y$ in (\ref{sys1}) represent the respective carrying capacities. Clearly, if $N_K, N_L, N_Y  \to \infty$, the system (\ref{sys1}) reduces to (\ref{sys}), meaning that the exponential model (\ref{sys}) is a limiting case of the logistic model (\ref{sys1}). Furthermore, it was shown in \cite{SW2024} that the two models (\ref{sys}) and (\ref{sys1}) are related via the following transformation: 
\begin{equation}
\label{tr}
\tilde{K} = \frac{N_K K}{N_K+K}, \quad \tilde{L} = \frac{N_L L}{N_L+L}, \quad \tilde{Y} = \frac{N_Y Y}{N_Y+Y}. 
\end{equation} 
For mathematical rigor and further details, see Section 6 in \cite{SW2024}. The transformation (\ref{tr}) effectively reduces the logistic model (\ref{sys1}) to the exponential model (\ref{sys}), demonstrating the connection between the two formulations. To derive the corresponding family of one-parameter production functions associated with the logistic model (\ref{sys1}), we solve for $K$, $L$, and $Y$   in (\ref{tr}), obtaining the inverse transformation given by
\begin{equation}
\label{tr1}
{K} = \frac{N_K \tilde{K}}{N_K-\tilde{K}}, \quad {L} = \frac{N_L \tilde{L}}{N_L-\tilde{L}}, \quad {Y} = \frac{N_Y \tilde{Y}}{N_Y-\tilde{Y}},  
\end{equation}
provided $\tilde{K} < N_K$, $\tilde{L} < N_L$, and $\tilde{Y} < N_Y$. Next, substituting the expressions for $K$, $L$, and $Y$ given by (\ref{tr1}) into (\ref{inv}) -- and similarly applying the transformation (\ref{tr1}) to the initial values $K_0$, $L_0$, and $Y_0$ --  we obtain, after dropping the tildes:
\begin{equation}
\label{inv2}
\begin{array}{l}
\displaystyle \left(\frac{N_K {K}}{N_K-{K}}\right)^{a_1}\left(\frac{N_L {L}}{N_L-{L}}\right)^{a_2}\left(\frac{N_Y {Y}}{N_Y-{Y}}\right)^{a_3} = \\ [0.5cm]
\displaystyle \left(\frac{N_K {K_0}}{N_K-{K_0}}\right)^{a_1}\left(\frac{N_L {L_0}}{N_L-{L_0}}\right)^{a_2}\left(\frac{N_Y {Y_0}}{N_Y-{Y_0}}\right)^{a_3} e^{(a_1b_1+a_2b_2+a_3b_3)t},
\end{array}
\end{equation}
where, as above,  $a_1,a_2, a_3 \in \mathbb{R}\setminus \{0\}$ are some parameters. The function  in three variables $K$, $L$, and $Y$ given by (\ref{inv2}) is a time-independent invariant of the flow generated by (\ref{sys1}) if and only if the orthogonality condition (\ref{con}) for the parameters $a_1$, $a_2$, and $a_3$ in (\ref{inv2}) holds true. Assuming the condition (\ref{con}) holds in (\ref{inv2}) and solving for $Y$, we arrive at the following expression: 
\begin{equation}
\label{nf}
Y = \frac{N_YBN_K^{-\frac{a_1}{a_3}}N_L^{-\frac{a_2}{a_3}}K^{-\frac{a_1}{a_3}}L^{-\frac{a_2}{a_3}}}{N_Y(N_K-K)^{-\frac{a_1}{a_3}}(N_L-L)^{-\frac{a_2}{a_3}} + BN_K^{-\frac{a_1}{a_3}}N_L^{-\frac{a_2}{a_3}}K^{-\frac{a_1}{a_3}}L^{-\frac{a_2}{a_3}}},
\end{equation} 
where 
$$B =  \left(\frac{N_K {K_0}}{N_K-{K_0}}\right)^{\frac{a_1}{a_3}}\left(\frac{N_L {L_0}}{N_L-{L_0}}\right)^{\frac{a_2}{a_3}}\left(\frac{N_Y {Y_0}}{N_Y-{Y_0}}\right). $$
Defining $\alpha$ and $\beta$ as before in (\ref{ab}) and rewriting the orthogonality condition (\ref{con}) accordingly -- see formula (\ref{con1}), we  derive from (\ref{nf}) the following formula: 
\begin{equation}
\label{nf1} 
Y= \frac{N_YK^{\alpha}L^{\frac{b_3}{b_2} - \alpha\frac{b_1}{b_2}}}{N_YB^{-1}N_K^{-\alpha}N_L^{-\frac{b_3}{b_2} + \alpha\frac{b_1}{b_2}}(N_K-K)^{\alpha}(N_L-L)^{\frac{b_3}{b_2} - \alpha\frac{b_1}{b_2}} + K^{\alpha}L^{\frac{b_3}{b_2} - \alpha\frac{b_1}{b_2}}},
\end{equation}
where 
\begin{equation}
\label{B}
B = \left(\frac{N_K {K_0}}{N_K-{K_0}}\right)^{-\alpha}\left(\frac{N_L {L_0}}{N_L-{L_0}}\right)^{-\frac{b_3}{b_2} + \alpha\frac{b_1}{b_2}}\left(\frac{N_Y {Y_0}}{N_Y-{Y_0}}\right). 
\end{equation} 
Clearly, the one-parameter family of the Cobb-Douglas functions given by (\ref{inv1}) is a limiting case of the one-parameter family of functions (\ref{nf1}) as $N_K$, $N_L$, and $N_Y$ $\to \infty$, as expected.  Therefore, we have proven the following 

\begin{proposition}{\em The logistic model (\ref{sys1}) admits a family  of production  functions (\ref{nf1}) as a time-independent invariant if and only if the orthogonality condition (\ref{con}) is satisfied.}
\end{proposition}
Next, we fix $\alpha$ in (\ref{nf1}), then the corresponding $\beta = \frac{b_3}{b_2} - \alpha\frac{b_1}{b_2}$ is also fixed. Extending the domain and range of the production function generalizing the Cobb-Douglas production function (\ref{nf1}) to $\mathbb{R}^2_+$ and $\mathbb{R}_+$ respectively, we arrive at

\begin{equation}
\label{3}
Y=\dfrac{N_{Y}L^\alpha K^\beta}{C|N_{L}-L|^\alpha |N_{K}-K|^\beta +L^\alpha K^\beta},     
\end{equation}
where as above $Y$ (production) is generated by $L$ (labor), $K$ (capital), while $\alpha$, $\beta$, $N_Y$, $N_K$,  $N_L$, and $$C =\left(\frac{N_K {K_0}}{N_K-{K_0}}\right)^{-\alpha}\left(\frac{N_L {L_0}}{N_L-{L_0}}\right)^{-\beta}\left(\frac{N_Y {Y_0}}{N_Y-{Y_0}}\right)$$  are positive constants. Importantly, the function (\ref{3}) has been empirically shown to provide a better fit than the Cobb-Douglas production function when compared to recent real-world data \cite{SW2019, SW2024}. Notably, it was first derived in \cite{SW2019} from the logistic model (\ref{sys1}) using integration.

It must be mentioned that another $S$-shaped production function, recently introduced as a generalization of the Cobb-Douglas function, has been studied in this context -- particularly within the framework of the Solow-Swan model -- leading to interesting results \cite{ACKL13, CET2012, EBC14, LLM15}. Specifically, this function is given by 
\begin{equation}
\label{nf3}
Y = \frac{A K^{p}L^{1-p}}{1 + BK^{p}L^{-p}},
\end{equation} 
where $A$, $B$, $p>0$ are constants and the variables $K$, $L$, and $Y$ denote capital, labor, and production respectively, as  before. We note that the production function (\ref{nf3}) exhibits constant returns to scale, which provides certain advantages in the development of economic growth models. However, to the best of our knowledge this function has neither been tested against economic data yet. Therefore, in what follows, we will employ the production function (\ref{3}) as a natural generalization of the Cobb-Douglas function (\ref{cd}) that can also be used in developing an extension of the Solow-Swan model.

Therefore, we built our model around the following assumptions: 

\begin{itemize}

\item[1.] There will be only one input --  energy ($E$). It is our contention that energy is the ultimate driving force behind production. In this view the production function we will build our model on will  be of the form $X = f(E)$. 

\item[2.]  Specifically, we will use the one-input version of the production function (\ref{3}) introduced and described for the first time in \cite{SW2019} defined as  
\begin{equation}
\label{4}
X = f(E)  =  \frac{N_1E^{\alpha}}{C|N_2- E|^{\alpha} + E^{\alpha}},
\end{equation}
where $X$ represents production, while $E$ denotes energy generation. Thereby, the  function (\ref{4}) serves as a foundational element of our model. 

\item[3.] We   assume that the value of produced energy depreciates over time at a constant rate $\delta$ and only a fraction of the output (production) $X(t)$  is used for energy production, that is the rate of change of $E$ as a function of time is equal to {\em the net energy gain (NEG)}, that is $$\dot{E} = {\mbox{Energy}}_{\mbox{\small consumable}} - {\mbox{Energy}}_{\mbox{\small expendable}}.$$ The dynamics of change of $E(t)$ is thus determined by the following differential equation:
\begin{equation}
\label{5}
\dot{E} = sX(E) - \delta E, \quad 0<s<1.    
\end{equation}
Note, that  (\ref{5}) is the key differential equation in our model, much like  (\ref{2}) -- in the Solow-Swan model. 
\end{itemize} 

The model aims to explain long-term economic growth by analyzing how energy production influences output over time. We aim to use this model to understand the role of energy production and its depreciation in economic growth. In this context, the steady state is the long-run equilibrium where energy production and output stabilize.

\section{Methods}

The renowned Cobb-Douglas production function \cite{CD1928} was generalized in \cite{SW2019} to the form (\ref{3}), incorporating logistic growth constraints on production, labor, and capital. This modified production function more accurately represents economic growth by preventing indefinite exponential expansion. Notably, \cite{SW2019} also introduced a single-input version of this function in a different context.

In this work, we employ this production function with energy production as its input. We generalize the Solow-Swan model \cite{RS1956} by replacing the Cobb-Douglas function with our proposed function (\ref{4}), where energy growth assumes the role traditionally held by capital accumulation. This approach yields an exogenous growth model that both builds upon established research and accounts for contemporary ecological and economic realities. Mathematically, we analyze this model using the same analytical framework applied to the Solow-Swan model and its generalizations: qualitative and quantitative theories of ordinary differential equations.

To fully comprehend the mathematical derivations in this paper, readers should be familiar with real analysis, dynamical systems theory, and integration techniques. We present all necessary methods prior to their application and include additional explanatory steps where warranted.

\section{Results}

In what follows we will thoroughly investigate the autonomous differential equation (\ref{5}) -- note the RHS of (\ref{5}) is independent of $t$. Next, we define the function 

$$F(E) := sX(E) - \delta E,$$
that is (\ref{5}) assumes the form $$\dot{E} = F(E).$$ First, we find the steady states, that is solve the equation $F=0$  for $E$. Clearly, $E_0=0$ is a steady state. To determine the other steady states, we solve the equation $F(E) = 0$, or
\begin{equation}
\frac{N_1E^{\alpha}}{C|N_2- E|^{\alpha} + E^{\alpha}} - \delta E = 0.
\label{6}
\end{equation}
After some algebra, we obtain
$$\delta C\Big|\frac{N_2}{E}-1\Big|^{\alpha} = \frac{N_1}{E} - \delta.$$

Since $E\not=0$, we introduce a new variable $x$ given by $x:=\frac{N_2}{E}-1$ and rewrite the above equation as 

\begin{equation}
\label{7}
\delta C|x|^{\alpha} = \frac{N_1}{N_2}x + \frac{N_1}{N_2} - \delta.
\end{equation}


The equation (\ref{7}) cannot be solved in full generality for an arbitrary $\alpha$. However, we can qualitatively analyze the existence of solutions. Indeed, note, $f(x) = \delta C|x|^{\alpha}$ is symmetric about the $y$-axis, while the linear function $g(x) = \frac{N_1}{N_2}x + \frac{N_1}{N_2} - \delta$ has the $x$-intercept in the interval $(-1,0)$, provided $\frac{N_1}{N_2} - \delta>0$.  This implies that the negative solution to the equation (\ref{6}) is greater than the $y$-intercept of $g(x)$ and greater than $-1$. Therefore, the corresponding value of $E = \frac{N_2}{1+x}$ is positive and so we have two positive solutions to (\ref{6}) in addition to $E_0=0$. The inequalities discussed above  provide the conditions of existence of non-trivial steady state solutions to~(\ref{6}). After some straightforward  stability analysis, it is clear to see that the greater non-trivial steady state, namely, $E_2$, is stable and the lesser non-trivial steady state, namely, $E_1$, is unstable. This agrees with the intuition that there is a threshold~(at $E_1$) below which energy plummets and above which energy grows. There exists, also, a greater threshold~(at $E_2$) above which energy once again diminishes, however, this time, to the non-zero steady state, $E_2$. Putting this together, we see that sufficient energy leads to sustainable growth which provides long term benefits to production.

To perform a quantitative analysis of the equation (\ref{7}), we will assume $\alpha=2$. This choice still captures the essence of the growth properties described by the formula (\ref{4}). Solving the equation $F(E)=0$ in this case, we arrive at

\begin{equation}
\label{8}
    E_0=0\text{, } E_{1,2}=\dfrac{sN_1+2c\delta N_2\mp\sqrt{4c\delta N_2(sN_1-\delta N_2)+s^2N_1^2}}{2\delta (c+1)}.
\end{equation}

It is clear to see from~(\ref{8}), that real solutions exist if and only if

\begin{equation}
1+\dfrac{sN_1}{4c\delta N_2}>\dfrac{\delta N_2}{sN_1}\text{.}
\end{equation}


Our model allows us to evaluate EROI (Energy Return on Investment), which is a measure of energy produced in relation to the energy used to create it. When the EROI  is high, it indicates that extracting energy from that source is relatively simple and cost-effective. Conversely, a low EROI suggests that obtaining energy from that source is challenging and costly. In the context of our model, the EROI is calculated as follows. 

\begin{equation}
\label{9}
    \mbox{EROI} := \dfrac{sX(E)}{\delta E}=\dfrac{sN_1E^\alpha}{\delta E(C|N_2-E|^\alpha +E^\alpha)}.
\end{equation}
Notably~(\ref{9}) relates energy expendature, of an amount $\delta E$, to produce $sX(E)$ energy. The formula (\ref{9}) for given fixed values of $N_1$, $N_2$, $\alpha$, $s$, and $\delta$ can be used to find the interval(s) where EROI is greater than a certain fixed quantity $r$: $\mbox{EROI} > r$. When $\alpha  = 2$ as above, it follows from (\ref{9}) that $\mbox{EROI} > r$ is a quadratic inequality that can be solved explicitly.

Our next goal is to produce a formula for the NEG (Net Energy Gain). We assume again that $\alpha=2$,   set without loss of generality $C=1$, and then solve the differential equation (\ref{5}) in this case. Replacing $E$ with some other variable $q$ for clarity, and integrating from 0 to $E$, we obtain the corresponding formula for NEG given by 

\begin{equation}\label{7a}
\begin{array}{rcl}
    \mbox{NEG} & := & \displaystyle \int_0^E f(q)dq\\[0.5cm]
    & = &\dfrac{sN_1\left(N_2\left(\ln\left|2\dfrac{E^2}{N_2^2}-2\dfrac{E}{N_2}+1\right|\right)+2E\right)-2\delta E^2}{4}\text{.}
\end{array}
\end{equation}
The formula~(\ref{7a}) provides a way to calculate the net gain or loss of energy as a system evolves from one state of energy to another. This provides insight into the short term intricacies resulting from an appreciating or depreciating economic system. Figure 1 shows an example of how NEG looks compared to energy growth and energy depreciation. It is important to note that a single use of~(\ref{7a}) provides only the net energy gain from the trivial steady state to an energy $E$. Note that we can analyze any movement within the system by utilizing~(\ref{7a}) two separate times. Suppose that the desired result is to analyze the NEG from energy of value $x$ to $y$. We first find the NEG from $0$ to $y$, then find the NEG from $0$ to $x$, finally we subtract the latter from the former and yield the desired result.

\begin{figure}
    \centering
    \includegraphics[width=0.75\textwidth]{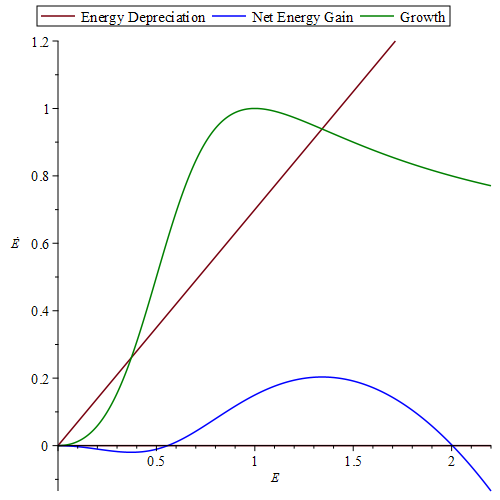}
    \caption{Stacked plot of energy growth (sX(t)), energy depreciation ($\delta E(t)$), and net energy gain (NEG). Parameters are as follows: $\alpha=2$, $s=1$ $C=1$, $N_1=1$, $N_2=1$, $\delta=0.7$.}
    \label{fig:1}
\end{figure}
We employ the following methods in order to produce a time-dependent solution in terms of energy as a function of time. Therefore, we  seek a solution of the form $E=f(t)$. To achieve this, we consider the following expression:

\begin{equation}
\dfrac{dE}{dt} = \dfrac{sN_1E^\alpha}{C(N_2-E)^\alpha +E^\alpha}-\delta E.
\end{equation}
For demonstration purposes we assign the following values in the above:  $s=1$, $C_1=1$, $N_1=1$, $N_2=1$, $\delta =0.7$, $\alpha =2$ and then evaluating the integral given by
\begin{equation}
\int \dfrac{(1-E)^2+E^2}{E^2-0.7E((1-E)^2+E^2)}dE=\int dt,
\end{equation}
resulting in the following implicit equation for a time-dependent energy solution: 
\begin{equation}
\label{10}
\dfrac{14E-12+\sqrt{46}}{14E-12-\sqrt{46}}E^{\left(-\dfrac{\sqrt{46}}{5}\right)} = \pm e^{\left(\dfrac{7\sqrt{46}}{50}(t+t_0)\right)}.
\end{equation}
\begin{figure}
    \centering
    \includegraphics[width=0.75\textwidth]{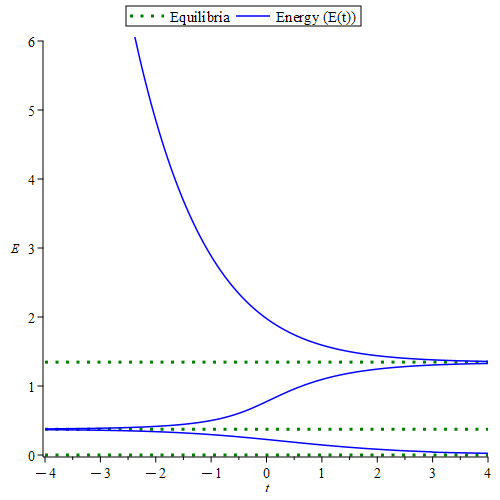}
    \caption{Energy as a function of time, dependent on initial energy as defined by (\ref{10}) along with energy steady states as defined by (\ref{8}) with the parameters   $s=1$, $C_1=1$, $N_1=1$, $N_2=1$, $\delta =0.7$, $\alpha =2$.}
    \label{fig:2}
\end{figure}
The equation~(\ref{10}) confirms the conclusion that there is a trivial steady state $E  = E_0$, which is stable, as well as two non-trivial steady states, one of which is stable ($E = E_2$) and the other ($E = E_1$) is unstable. To produce the graph of the function $E  = E(t)$, we generate the graph $t=t(E)$ from (\ref{10}) first and then reflect it about the line $E = t$, since $E= E(t)$  is the inverse function of $t = t(E)$.

Figure \ref{fig:2} illustrates the temporal evolution of energy under various initial conditions. The solutions to (\ref{10}) exhibit three distinct behavioral regimes:
\begin{enumerate}

\item[1.] {\em Sustainable Growth}: Solutions with initial conditions above $E_1$ converge to long-term sustainable growth.

\item[2.] {\em Growth Collapse}: Solutions starting below $E_1$ fail to sustain long-term growth.

\item[3.] {\em Transition Dynamics}:

\begin{itemize}
\item[--] Initial conditions above $E_2$ exhibit short-term energy loss due to excessive expenditure.

\item[--] Initial conditions between $E_1$ and $E_2$ represent the optimal regime, balancing short-term growth with long-term sustainability as they asymptotically approach $E = E_2$.
\end{itemize}
\end{enumerate}
\section{Discussion}

Environmental sustainability has emerged as a critical consideration in contemporary economic discourse. Recognizing this imperative, we present a growth model that reconciles classical economic theory with modern ecological constraints. By adopting the framework developed in \cite{SW2019}, we utilize a production function that better aligns with current empirical evidence while incorporating essential environmental dimensions. Our approach integrates critical energy considerations by adapting an established single-input growth model to explicitly capture energy dynamics (\ref{4}). This methodological innovation extends recent advances in factor-specific growth modeling while introducing essential environmental dimensions to the study of economic expansion. By doing so, we bridge a persistent gap in growth theory, offering new analytical tools to examine the energy-growth nexus within an ecological framework.

Building on the foundational Solow-Swan model, our framework provides policymakers with a robust analytical tool for sustainable economic planning. The model achieves this through three key innovations:

\begin{enumerate}
\item[1.] Optimization of net energy gain (\ref{7}).

\item[2.] Projection of growth trajectories via time-dependent solutions (\ref{10}).

\item[3.] Integrated stability and EROI analysis (\ref{9}).
\end{enumerate}

These components work synergistically to inform investment strategies that reconcile immediate economic needs with long-term ecological constraints. By combining quantitative rigor with qualitative insights, our approach offers a practical pathway to sustainable, inclusive growth.



\end{document}